\documentclass[cits]{PoS}

\usepackage{amsmath}

\newcommand{\gev}{\,\operatorname{GeV}}

\newcommand{\fm}{\operatorname{fm}}

\newcommand{\mb}{\operatorname{mb}}

\newcommand{\ms}{\mskip 1.5mu}

\newcommand{\jpsi}{J\mskip -2mu/\mskip -0.5mu\Psi}

\renewcommand{\vec}[1]{\textbf{#1}}

\title{\raggedleft{\small DESY 13-112}  \\[2em] \raggedright
  Correlation effects in multiple hard scattering}

\ShortTitle{Correlation effects in multiple hard scattering}

\author{\speaker{Markus DIEHL}\\
  Deutsches Elektronen-Synchroton DESY, 22603 Hamburg, Germany \\
  E-mail: \email{markus.diehl@desy.de}}

\abstract{Correlations between the incoming partons in multiple hard
  scattering can affect both the event rate and kinematic distributions in
  the final state.  In this talk, I discuss different types of
  correlations relevant for double parton scattering in proton-proton
  collisions.}

\FullConference{XXI International Workshop on Deep-Inelastic Scattering
  and Related Subjects-DIS2013,\\
  22-26 April 2013\\
  Marseilles,France}

\begin{document}

\section{Introduction}

Multiparton interactions in proton-proton collisions, i.e.\ interactions
in which more than one parton in one proton scatters on partons in the
other proton, pose a challenge for understanding the hadronic final state
at the LHC.  In this talk, I concentrate on the case with two
hard-scattering subprocesses in one collision, termed double parton
scattering (DPS).  Assuming factorization, the leading-order cross section
for DPS can be written as
\begin{align}
  \label{dps-fact}
 \frac{d\sigma_{\text{double}}}{dx_1\, d\bar{x}_1\;
  dx_2\, d\bar{x}_2 \rule{0pt}{1.8ex}}
&= \frac{1}{C}\, \sum_{a_1 a_2 \ms \bar{a}_1 \bar{a}_2 \rule{0pt}{1.2ex}}
\hat{\sigma}_{a_1 \bar{a}_1}\,
\hat{\sigma}_{a_2 \bar{a}_2}
\int\! d^2\vec{b}\,\,
F_{a_1 a_2}(x_1, x_2, \vec{b}) \,
F_{\bar{a}_1 \bar{a}_2}(\bar{x}_1, \bar{x}_2, \vec{b}) \,,
\end{align}
where $\hat{\sigma}_{a_i \bar{a}_i}$ is the cross section for the hard
subprocess initiated by partons $a_i$ and $\bar{a}_i$, and $C$ is a
combinatorial factor.  $F_{a_1 a_2}(x_1, x_2, \vec{b})$ is a double parton
distribution (DPD), which depends on the momentum fractions $x_i$ of the
two partons and on their relative transverse distance $\vec{b}$.
Higher-order corrections to $\hat{\sigma}_{a_i \bar{a}_i}$ can be
incorporated into \eqref{dps-fact} in the same way as for single hard
scattering.
DPDs are essentially unknown functions, and for phenomenology one needs an
ansatz for them.  The simplest assumption is to write them as a product
\begin{align}
\label{dpd-factor}
F_{a_1 a_2}(x_1,x_2,\vec{b}) 
 &= f_{a_1}(x_1)\, f_{a_2}(x_2)\, G(\vec{b})
\end{align}
of single-parton densities $f_{a}(x)$ and a factor $G(\vec{b})$ describing
the dependence on the interparton distance.  Assuming that this factor is
the same for all parton types, we obtain
\begin{align}
  \label{pocket-form}
 \frac{d\sigma_{\text{double}}}{dx_1\, d\bar{x}_1\;
  dx_2\, d\bar{x}_2 \rule{0pt}{1.8ex}}
&= \frac{1}{C}\,
\frac{d\sigma_1}{dx_1\, d\bar{x}_1 \rule{0pt}{1.8ex}} \;
\frac{d\sigma_2}{dx_2\, d\bar{x}_2 \rule{0pt}{1.8ex}}\;
       \frac{1}{\sigma_{\text{eff}}}
\end{align}
from \eqref{dps-fact}, with a universal factor $1 /\sigma_{\text{eff}} =
\textstyle{\int} d^2\vec{b}\; \bigl[ G(\vec{b}) \bigr]{}_{}^2$ and cross
sections $\sigma_{i}$ for single hard scattering.  This ``pocket formula''
provides a convenient first estimate for the size of DPS in different
processes.  Experimental studies at the Tevatron and the LHC yield values
of $\sigma_{\text{eff}}$ between $10$ and $20 \mb$, which are consistent with
each other within their errors \cite{{Diehl:DISplen}}.

Let us see what follows for $\sigma_{\text{eff}}$ if there are no
correlations at all in the distribution of two partons.  The DPD can then
be written as
\begin{align}
  \label{dpd-indept}
F_{a_1 a_2}(x_1,x_2,\vec{b}) = {\int} d^2\vec{b}'\;
  f_{a_1}(x_1,\vec{b}'+\vec{b})\, f_{a_2}(x_2,\vec{b}') \,,
\end{align}
where $f_a(x, \vec{b})$ is the probability density for finding a parton
$a$ with momentum fraction $x$ at transverse position $\vec{b}$ inside the
proton.  Information about this distribution can be inferred from
generalized parton distributions measured in exclusive scattering
processes \cite{Sabatie:DIS2013} and from nucleon form factors
\cite{Diehl:DISspin}.  If one further assumes a universal $\vec{b}$
dependence for all partons, $f_{a}(x, \vec{b}) = f_{a}(x)\, F(\vec{b})$,
then the DPD factorizes as in \eqref{dpd-factor}, with
\begin{align}
  \label{G-factored}
G(\vec{b}) 
 &= {\int} d^2\vec{b}'\; F(\vec{b}'+\vec{b})\, F(\vec{b}') \,.
\end{align}
For a Gaussian dependence of $F(\vec{b})$ with average $\langle \vec{b}^2
\rangle$ one obtains
\begin{align}
  \label{sigma-eff-indpt}
\sigma_{\text{eff}} = 4\pi \langle \vec{b}^2 \rangle
  = 41 \mb \; \times
    \frac{\langle \vec{b}^2 \rangle}{(0.57 \fm)^2} \,.
\end{align}
If $F(\vec{b})$ is the Fourier transform of a dipole $1 / (1 + \Delta^2
/M^{\,2})^{2}$ in transverse momentum space, then there is an additional
factor of $7/8$ on the r.h.s., so that the prefactor reads $36 \mb$
instead of $41 \mb$.  Studies of generalized parton distributions give a
typical range $\langle \vec{b}^2 \rangle \sim (0.57 \fm -\, 0.67 \fm)^2$.
The resulting values of $\sigma_{\text{eff}}$ are hence significantly
larger than those extracted from DPS processes.  This mismatch has been
noticed long ago \cite{Calucci:1997ii,Frankfurt:2003td}, and different
possible reasons have been put forward.  Quite obviously, correlations in
the distribution of two partons
\cite{Calucci:1997ii,Frankfurt:2003td,Calucci:1999yz,Calucci:2009sv,%
  Calucci:2010wg,Frankfurt:2004kn,Rogers:2009ke,Domdey:2009bg,%
  Flensburg:2011kj,Blok:2012mw,Blok:2013bpa} could invalidate the estimate
based on \eqref{dpd-indept}.  Such correlations can be of quite different
type, and in the following we will discuss some of them in more detail.

Notice that there could be correlations between two partons that
invalidate the form \eqref{dpd-indept} but still preserve factorization of
the type \eqref{dpd-factor}.  The pocket formula \eqref{pocket-form} would
then hold, but $\sigma_{\text{eff}}$ could no longer be calculated from
the one-particle distribution $f_a(x,\vec{b})$.  We will, however, see in
the following that it is plausible to expect correlations that invalidate
\eqref{dpd-factor}.  In this case, sufficiently precise extractions of
$\sigma_{\text{eff}}$ based on \eqref{pocket-form} will find a dependence
of this quantity on the process or on kinematics.


\section{Correlations between $x_1$ and $x_2$}

It is clear that the form \eqref{dpd-indept} cannot hold when $x_1$ or
$x_2$ becomes large, if only because DPDs should vanish at the kinematic
boundary $x_1+x_2 = 1$.  To alleviate this problem, a modified ansatz
$F_{a_1 a_2}(x_1,x_2,\vec{b}) = f_{a_1}(x_1)\, f_{a_2}(x_2)\,
(1-x_1-x_2)^n\, G(\vec{b})$ with some power $n$ is sometimes used to
suppress the region of large $x_1+x_2$.
Significant correlations between the $x_1$ and $x_2$ dependence of
$F(x_1,x_2,\vec{b})$ have been found in a recent calculation using a
constituent quark model \cite{Rinaldi:2013vpa}, which should give a first
idea on the behavior of DPDs for $x_1$ and $x_2$ above, say, $0.1$.

Both considerations do, however, leave open the possibility that
correlations between the momentum fractions of the two partons are weak at
small $x_1$ and $x_2$, which is the kinematic region relevant for most
processes at the LHC.


\section{Correlations between $x_1, x_2$ and $\vec{b}$}

Studies of hard exclusive processes and of the nucleon form factors have
taught us about generalized parton distributions and thus, with some
degree of model dependence, about the impact parameter distribution
$f_a(x,\vec{b})$ of a single parton inside the proton.  In particular,
measurements of $\gamma p\to \jpsi p$ at HERA
\cite{Chekanov:2002xi,Aktas:2005xu} indicate a weak logarithmic dependence
$\langle \vec{b}^2 \rangle = \text{const} + 4 \alpha' \log(1/x)$ with
$\alpha' \approx 0.15 \gev^{-2} = (0.08 \fm)^2$ for gluons with $x$ around
$10^{-3}$.  Studies of nucleon form factors \cite{Diehl:DISspin} and
calculations of Mellin moments $\int dx\; x^n f_a(x,\vec{b})$ with
$n=0,1,2$ in lattice QCD \cite{Hagler:2009ni} indicate that for $x$ above
$0.1$ the decrease of $\langle \vec{b}^2 \rangle$ with $x$ is even
stronger.  Future measurements at JLab, COMPASS and (hopefully one day) at
EIC and LHeC will yield a detailed quantitative picture of the
one-particle distributions $f_a(x,\vec{b})$ in the nucleon.

Given that the $\vec{b}$ distribution of a single parton becomes more
narrow with increasing $x$, it is plausible to assume a similar
correlation between the $\vec{b}$ dependence and $x_1, x_2$ in double
parton distributions, even if the ansatz \eqref{dpd-indept} of independent
partons does not hold.  As pointed out in \cite{Frankfurt:2003td}, this
has important consequences for multiparton interactions.  The production
of a particles with high transverse momentum or mass requires relatively
large momentum fractions of the partons entering the corresponding hard
interaction.  This favors smaller values of $\vec{b}$ and thus a higher
transverse overlap between the colliding protons, which in turn favors
further interactions.  An study with Pythia 8 has shown that such
correlations have a quantitative impact for instance on the underlying
event activity in $Z$ production \cite{Corke:2011yy}.  Further
investigation along these lines would be very interesting.


\section{Spin correlations}

Even in an unpolarized proton, the polarizations of two quarks or gluons
can be correlated with each other.  As was already noted in
\cite{Mekhfi:1985dv}, such correlations have measurable consequences for
DPS processes.  Spin correlations between the incoming partons can be
encoded in polarized DPDs, which need to be included in the factorization
formula \eqref{dps-fact}.  If they are sizeable, these extra terms will
invalidate the derivation leading to the pocket formula
\eqref{pocket-form}.

A full classification of polarized double parton distributions for quarks
and gluons has been given in \cite{Diehl:2011yj,Diehl:2013mla}, and in
\cite{Diehl:2013mla} it was shown that their size is limited by positivity
bounds similar to the Soffer bound \cite{Soffer:1994ww} for polarized
single-parton densities.
The implications of parton spin correlations have been investigated in
\cite{Kasemets:2012pr} for the production of two electroweak gauge bosons
($\gamma^*, Z, W$) followed by their decay into leptons.  One finds that
longitudinal quark spin correlations affect both the rate of DPS and the
rapidity distribution of the decay leptons.  By contrast, transverse
quark spin correlations lead to a modulation in the azimuthal angle
between the lepton decay planes of the two bosons.  This makes it most
evident that the two hard-scattering processes in DPS are not completely
independent if there are correlations between the incoming partons.

A study in the MIT bag model found very significant spin correlations in
DPDs \cite{Chang:2012nw}, and one can expect a similar trend in any model
that describes the proton in terms of three quarks.  The perturbative
splitting of a single parton into two, which describes the behavior of DPDs
at small interparton distance $\vec{b}$, also generates strong spin
correlations \cite{Diehl:2011yj}.  How important spin correlations are at
small $x_i$ and large $\vec{b}$ remains currently unknown.


\section{Color correlations}

Not only the spin but also the color of two partons can be correlated, and
as a consequence, further terms with DPDs describing color correlations
need to be added in the factorization formula \eqref{dps-fact}.  In the
familiar collinear factorization formalism, such color correlations are
suppressed by Sudakov factors \cite{Mekhfi:1988kj,Manohar:2012jr}.  The
amount of suppression depends on the hard scale of the process, and an
estimate in \cite{Manohar:2012jr} finds a suppression factor that is very
small for $Q = 100 \gev$, but only equal to 0.5 for $Q=10 \gev$.

For small measured transverse momenta, the discussion of Sudakov factors
is more involved.  If the produced particles are color neutral, one can
extend the Collins-Soper-Sterman analysis of single Drell-Yan production
\cite{Collins:1984kg} to DPS as shown in \cite{Diehl:2011yj}.  In this
context, color correlation effects become relevant beyond the leading
double logarithmic accuracy.


\section{Conclusions}

A variety of correlations between two partons can affect the overall rate
of double parton scattering processes, their kinematic dependence, and
differential distributions in the final state.  Moreover, such
correlations are a characteristic feature of proton structure and thus of
interest in their own right.  There are indirect hints for correlations of
non-negligible size, but there is no unambiguous evidence for far.
Further theoretical and experimental studies in this area are definitely
needed.


\end{document}